\documentclass[twocolumn,english,superscriptaddress,floatfix]{revtex4}
\usepackage[T1]{fontenc}
\usepackage[latin9]{inputenc}
\usepackage{color}
\usepackage{babel}
\usepackage{amsmath}
\usepackage{amssymb}
\usepackage{graphicx}
\usepackage{esint}
\usepackage[unicode=true,
 bookmarks=false,
 breaklinks=true,pdfborder={0 0 1},backref=section,colorlinks=true]
 {hyperref}
\hypersetup{
 pdfcreator={},pdfproducer={LaTeX with hyperref},linkcolor=blue,anchorcolor=blue,citecolor=blue,filecolor=red,menucolor=red,pagecolor=red,urlcolor=blue,pdfstartview=FitV,pdfhighlight=/I,pdfpagelayout=OneColumn,hypertexnames=true}

\makeatletter
\@ifundefined{textcolor}{}
{%
 \definecolor{BLACK}{gray}{0}
 \definecolor{WHITE}{gray}{1}
 \definecolor{RED}{rgb}{1,0,0}
 \definecolor{GREEN}{rgb}{0,1,0}
 \definecolor{BLUE}{rgb}{0,0,1}
 \definecolor{CYAN}{cmyk}{1,0,0,0}
 \definecolor{MAGENTA}{cmyk}{0,1,0,0}
 \definecolor{YELLOW}{cmyk}{0,0,1,0}
}

\usepackage{babel}

\usepackage{upgreek}

 \@ifundefined{textcolor}{}{%
  \definecolor{BLACK}{gray}{0}
  \definecolor{WHITE}{gray}{1}
  \definecolor{RED}{rgb}{1,0,0}
  \definecolor{GREEN}{rgb}{0,1,0}
  \definecolor{BLUE}{rgb}{0,0,1}
  \definecolor{CYAN}{cmyk}{1,0,0,0}
  \definecolor{MAGENTA}{cmyk}{0,1,0,0}
  \definecolor{YELLOW}{cmyk}{0,0,1,0}
 }

 \@ifundefined{textcolor}{}{
  \definecolor{BLACK}{gray}{0}
  \definecolor{WHITE}{gray}{1}
  \definecolor{RED}{rgb}{1,0,0}
  \definecolor{GREEN}{rgb}{0,1,0}
  \definecolor{BLUE}{rgb}{0,0,1}
  \definecolor{CYAN}{cmyk}{1,0,0,0}
  \definecolor{MAGENTA}{cmyk}{0,1,0,0}
  \definecolor{YELLOW}{cmyk}{0,0,1,0}
 }
 \@ifundefined{definecolor}{
 }{}
 \@ifundefined{definecolor}{color}{}

\usepackage{color}

\newcommand{\be}{\begin{equation}}
\newcommand{\ee}{\end{equation}}
\newcommand{\bea}{\begin{eqnarray}}
\newcommand{\eea}{\end{eqnarray}}
\newcommand{\bse}{\begin{subequations}}
\newcommand{\ese}{\end{subequations}}

\usepackage{color}
\definecolor{d_red}{cmyk}{0.00, 0.81, 1.00, 0.27}
\definecolor{d_orange}{cmyk}{0.00, 0.33, 1.00, 0.00}
\definecolor{d_blue}{cmyk}{0.78, 0.47, 0.00, 0.20}
\definecolor{d_lgreen}{cmyk}{0.07, 0.00, 0.79, 0.29}
\definecolor{d_green}{cmyk}{0.66, 0.00, 0.71, 0.56}
\definecolor{d_blue}{cmyk}{0.78, 0.47, 0.00, 0.20}
\definecolor{d_dblue}{cmyk}{0.91, 0.79, 0.00, 0.22}
\definecolor{d_pink}{cmyk}{0.0, 0.79, 0.37, 0.29}
\definecolor{d_purple}{cmyk}{0.16, 0.54, 0.00, 0.70}
\definecolor{d_paleblue}{cmyk}{0.669, 0.338, 0.00, 0.373}
\definecolor{d_dpaleblue}{cmyk}{0.441, 0.290, 0.00, 0.580}
\definecolor{d_brown}{cmyk}{0.0, 0.490, 0.930, 0.350}
\definecolor{d_turquoise}{cmyk}{0.630, 0.04, 0.0, 0.440}
\definecolor{KIT-green}{RGB}{0, 150,130}
\definecolor{KIT-blue}{RGB}{70,100,170}

\def\bmx{\begin{pmatrix}}
\def\emx{\end{pmatrix}}

\usepackage[figure,table]{hypcap}

\makeatother

\begin{document}
\title{Quantum discontinuity fixed point and renormalization group flow of
the SYK model }
\author{Roman Smit}
\affiliation{Institut für Theoretische Physik, Universität Frankfurt, Max-von-Laue
Strasse 1, 60438 Frankfurt, Germany}
\author{Davide Valentinis}
\affiliation{Institut für Quantenmaterialien und Technologien, Karlsruher Institut
für Technologie, 76131 Karlsruhe, Germany}
\author{Jörg Schmalian}
\affiliation{Institut für Theorie der Kondensierten Materie, Karlsruher Institut
für Technologie, 76131 Karlsruhe, Germany}
\affiliation{Institut für Quantenmaterialien und Technologien, Karlsruher Institut
für Technologie, 76131 Karlsruhe, Germany}
\author{Peter Kopietz}
\affiliation{Institut für Theoretische Physik, Universität Frankfurt, Max-von-Laue
Strasse 1, 60438 Frankfurt, Germany}
\begin{abstract}
We determine the global renormalization group (RG) flow of the Sachdev-Ye-Kitaev
(SYK) model. This flow allows for an understanding of the surprising
role of critical slowing down at a quantum first-order transition
in strongly-correlated electronic systems. From a simple truncation
of the infinite hierarchy of the exact functional RG flow equations
we identify several fixed points: Apart from a stable fixed point, associated with the celebrated non-Fermi
liquid state of the model, we find another stable fixed point related to an
integer-valence state. These stable fixed points
are separated by a discontinuity fixed point with one relevant direction, describing a quantum first-order transition.   Most notably, the fermionic
spectrum continues to be quantum critical even at the discontinuity
fixed point. This rules out a description of this quantum first-order
transition in terms of a local effective Ising variable that is established
for classical transitions. It reveals that quantum phase coexistence can be a genuine critical state of matter.
\end{abstract}

\date{October 2, 2020}

\maketitle

The liquid-vapor transition of real gases is the prime example of
a first-order phase transition that preserves the symmetry and terminates
at a critical end-point~\cite{vdWaals73}. Such transitions are of
importance in systems as diverse as hot and dense nuclear matter~\cite{Stephanov2004},
polymer-gel fluid mixtures~\cite{Tanaka78}, and correlated-electron
systems. The famous Mott transition between states of localized and
delocalized electrons~\cite{Mott90,Castellani79,Kotliar00,Majumdar94,Limelette03,Hassan05,Lefebvre00,Kagawa05,deSouza05,Bartosch10,Gati16,Pustogow18,Pustogow19,Rosslhuber20},
the Kondo volume-collapse transition~\cite{Lawrence81,Sarrao99,Allen82,Dzero2010,Lanata13},
or valence transitions in inter-metallic compounds~\cite{Honda16}
are prominent correlated-electron problems of this kind. Significant
experimental and theoretical insights have been gained with regards
to the behavior of the classical critical end-point of these electronic
states; it is described in terms of an effective Ising model and governed
by critical elasticity, a vanishing bulk modulus, and rich crossover
behavior due to the finite shear modulus of crystalline solids~\cite{Castellani79,Kotliar00,Levanyuk70,Hackl08,Papanikolaou08,Zacharias12,Zacharias15}. 

\begin{figure}[tb]
\begin{centering}
 \includegraphics[width=0.5\textwidth]{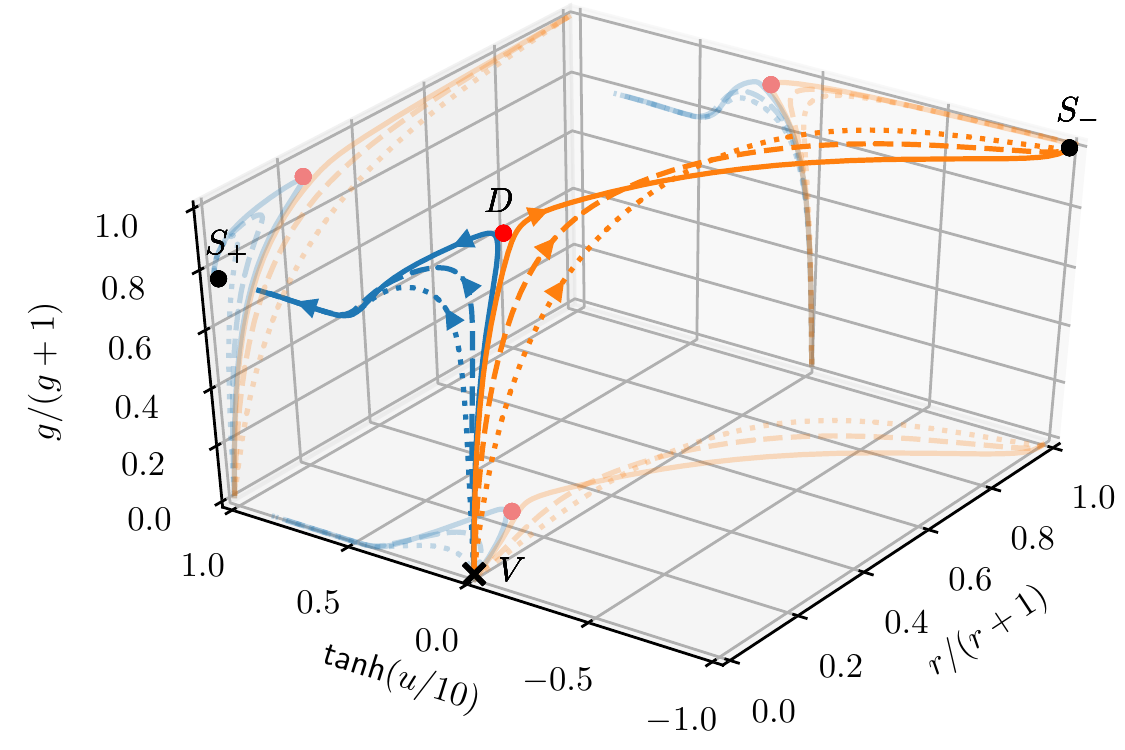}
\par\end{centering}
\caption{Global RG flow of the SYK model from Eq.~(\ref{eq:gflow}). The dimensionless coupling
$r$ determines the flow of $\Sigma\left(0\right)-\mu$ (where $\Sigma (0)$ is the 
self-energy at frequency $\omega=0$ and $\mu$ is the chemical potential) 
while $u$ and $g$ are rescaled four and six-point vertices for vanishing frequencies. At high
energies, the flow starts near the vacuum fixed point $V$. The discontinuity
fixed point $D$ separates the stable fixed points $S_{+}$ and $S_{-}$
characterized by non-Fermi liquid and integer-valence behavior, respectively.
Most notably, the first-order transition described by $D$ is characterized
by quantum-critical fermions with anomalous dimension $\eta=1/2$.}
\label{fig:flow3d}
\end{figure}

Much less is known about the associated quantum first-order transition.
It is an open question whether the description in terms of an effective
Ising model, successful for classical transitions, continues to be
appropriate. This is particularly ambiguous if the transition is from
a non-Fermi liquid, strange-metal state to a fully localized incompressible
state of matter. Exotic behavior was found near several quantum first-order
transitions~\cite{Pfleiderer2005,Joerg2010,Laumann2012,Zhao2019}.
A symmetry-preserving first-order transition that terminates at a
critical end-point was recently identified~\cite{Banerjee2017,Azeyanagi2018,Patel2019,Ferrari19,Wang2020,Sorokhaibam20}
in the Sachdev-Ye-Kitaev (SYK) model~\cite{Sachdev93,Kitaev15} and
generalizations thereof. The SYK model proved important for our understanding
of the intriguing properties of interacting quantum matter without
quasiparticles~\cite{Patel2019,Sachdev15,Polchinski16,Maldacena16,Song17,Rosenhaus18,Hartnoll18,Trunin20,Esterlis19,Wang20}.
The ability to perform controlled calculations in the strong-coupling
regime of the SYK model makes it a promising platform to elucidate
the role of electronic dynamics at a quantum first-order transition.

In this paper, using a functional renormalization group (FRG) approach,
we analyze the scaling behavior at a first-order quantum phase transition
in the SYK model. Performing a large-$N$ truncation of the formally
exact FRG flow equations \cite{Salmhofer01,Kopietz01,Kopietz10,Metzner12,Dupuis20},
we derive the global renormalization group (RG) flow of the SYK model at zero temperature. As
shown in Fig.\ref{fig:flow3d}, a flow profile emerges that exhibits
four different fixed points: Apart from a trivial vacuum fixed point
$V$ we find two stable fixed points $S_{+}$ and $S_-$ separated by a
discontinuity fixed point $D$. The latter is driven by changing the
chemical potential $\mu$ and, as we will show, describes the first-order transition
between a critical non-Fermi liquid state ($S_{+}$) and an integer-valence
state ($S_{-}$). The equation of state of this transition was analyzed
in Refs.~\cite{Azeyanagi2018,Ferrari19}; similar transitions were discussed
in a number of closely related models~\cite{Banerjee2017,Patel2019,Wang2020},
reflecting for example a drastic change from fast to slow quantum
information scrambling~\cite{Banerjee2017}. Here we determine the
relevant exponent of the discontinuity fixed point in accordance with
the scaling theory of first-order transitions~\cite{Nienhuis75,Fisher82}.
In addition, we find that fermions at the transition are governed
by quantum-critical dynamics with an anomalous dimension, despite
the discontinuous change of thermodynamic variables. Our results are
a consequence of the subtle interplay between high-energy and
low-energy dynamics that can be captured by the FRG approach developed here. 
Our results may also shed light onto the recent observation of powerlaw behavior of the dielectric function at a first-order Mott transition~\cite{Pustogow19,Rosslhuber20}.

\begin{figure}[tb]
\begin{centering}
\centering
 \vspace{5mm}
 \includegraphics[width=0.44\textwidth]{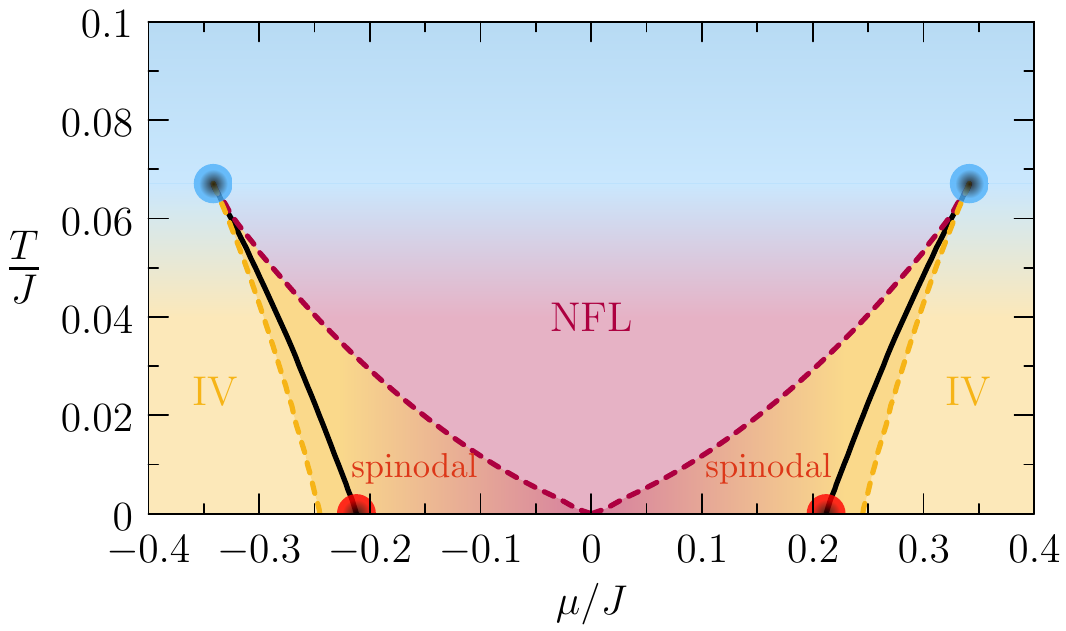}
\par\end{centering}
\caption{Phase diagram of the SYK model as a function of chemical potential
$\mu$ and temperature $T$ obtained from the numerical solution of
the large-$N$ self-consistency equation for the self-energy; see
also Ref.~\cite{Azeyanagi2018}. For $T=0$ and $|\mu|\protect\leq\mu_{\ast}=0.212J$,
the non-Fermi liquid (NFL) phase with anomalous dimension $\eta=1/2$
is the stable solution. For $\mu=\pm\mu_{\ast}$ (red dots) there are first order
quantum phase transitions from  $n\approx 0.76$ or $0.24$ to integer-valence (IV) phases $n=1$ or $0$, respectively. Between the
spinodal lines (dashed) both phases are locally stable. The transition
terminates at critical points $(\mu_{c},T_{c})=(\pm0.34,0.067)J$ (blue dots).}
\label{fig:phasediagram}
\end{figure}

\emph{SYK model:} 
The SYK model describes $N$ species of fermions
interacting with random two-body interactions, 
\begin{eqnarray}
{\cal H} & =-\mu\sum_{i}c_{i}^{\dagger}c_{i}+ & \sum_{i<j,k<l}J_{ij,kl}c_{i}^{\dagger}c_{j}^{\dagger}c_{k}c_{l}.\label{eq:Hdef}
\end{eqnarray}
Here
$c_{i}$ and $c_{i}^{\dagger}$ are fermionic annihilation and creation
operators, $i,j\cdots$ label $N$ different orbitals or lattice sites, and
$\mu$ is the chemical potential. The $J_{ij,kl}$ are random variables
with Gaussian distribution of zero mean and variance $\overline{\left|J_{ij,kl}\right|^{2}}=2J^{2}/N^{3}$.
The model is exactly solvable in the limit $N\rightarrow\infty$ where
the single-particle properties are determined by the self energy $\Sigma\left(\tau\right)=-J^{2}G^{2}\left(\tau\right)G\left(-\tau\right)$
together with the Dyson equation $G^{-1}\left(\omega\right)=i\omega+\mu-\Sigma\left(\omega\right)$.
At temperature $T=0$ one can construct two solutions of this set of equations.
On the one hand, there is the critical, non-Fermi liquid (NFL) solution
which takes for  $\left|\omega\right|\ll J$ the form
\begin{equation}
\Sigma_{{\rm NFL}}\left(\omega\right)=\mu-\frac{JA\left(\theta,\omega\right)}{\pi^{1/4}}\left|\frac{\omega}{J}\right|^{\eta},\label{eq:NFLsol}
\end{equation}
with anomalous dimension $\eta=1/2$~\cite{Delta}. The coefficient
$A=\cos^{1/4}\left(2\theta\right)e^{-i{\rm sign}\left(\omega\right)\left(\theta-\frac{\pi}{2}\right)}$
depends on the particle number $n=1/2+\theta/\pi+\sin\left(2\theta\right)/4$
through the angle $\theta\in\left[-\frac{\pi}{4},\frac{\pi}{4}\right]$~\cite{Sachdev15}.
The solution, Eq.(\ref{eq:NFLsol}), with power-law propagator $G\left(\tau\right)\propto\tau^{-1/2}$
yields the much discussed finite ground state entropy, displays analogies
to the physics of black holes with regards to its information scrambling,
led to theories of the transport properties of strange metals, and
allowed for the analysis of superconductivity in non-Fermi liquids~\cite{Patel2019,Sachdev15,Polchinski16,Maldacena16,Song17,Rosenhaus18,Hartnoll18,Trunin20,Esterlis19,Wang20}.
However, at $T=0$ and finite $\mu$ one can easily show that
\begin{equation}
\Sigma_{{\rm IV}}\left(\omega\right)=0\label{eq:IVsol}
\end{equation}
is another solution of the large-$N$ equations describing an integer-valence
(IV) state. Depending on the sign of $\mu$ we have $n=1$ or $0$,
i.e. a completely filled or empty system. As shown in Ref.~\cite{Azeyanagi2018}
a first-order transition occurs between both solutions at $\mu=\pm\mu_{*}\left(T\right)$.
\begin{figure}[tb]
\begin{centering}
\includegraphics[width=0.29\textwidth]{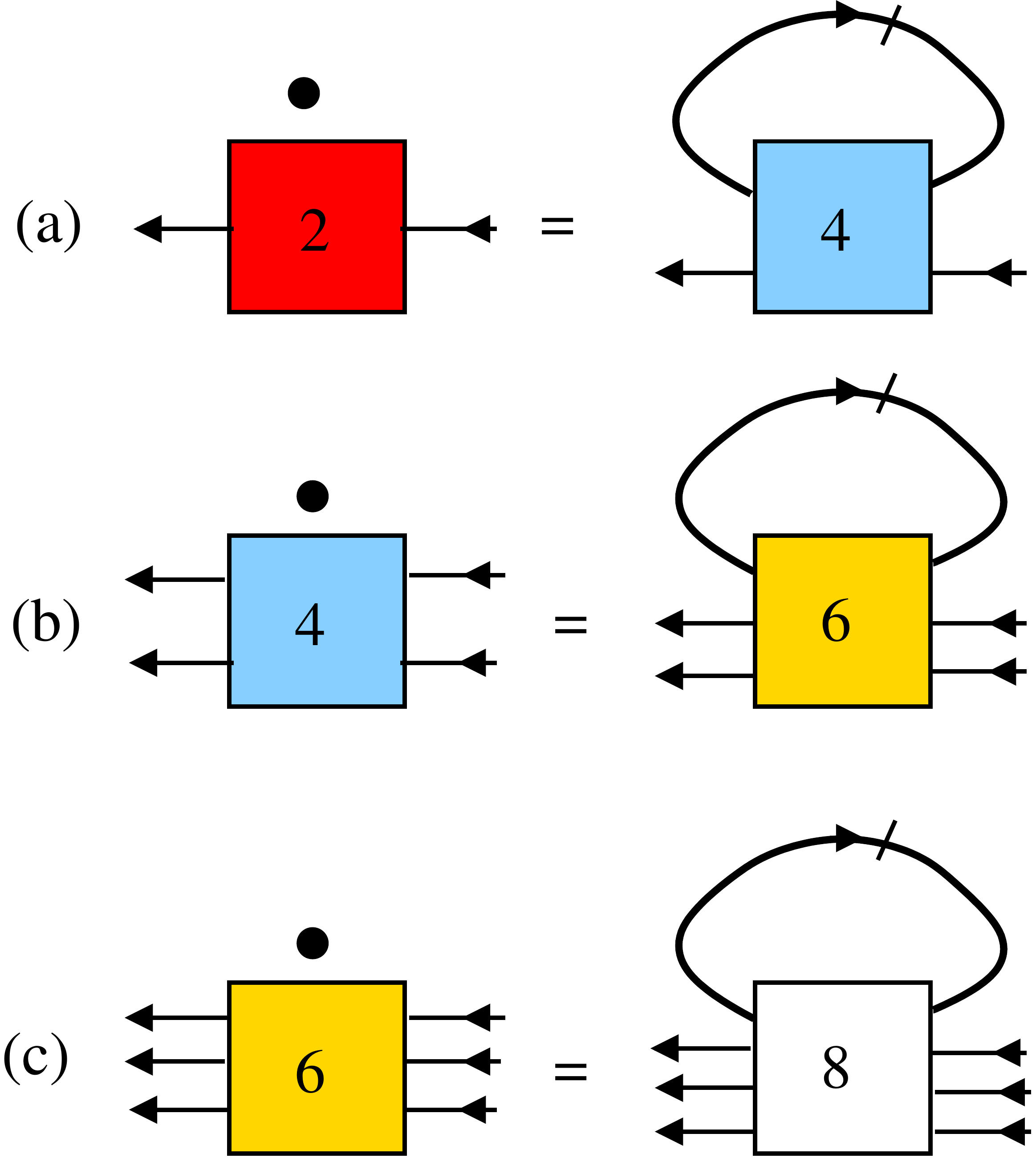}
\par\end{centering}
\caption{Graphical representation of the truncated hierarchy of FRG flow equations
leading to Eq.~(\ref{eq:gflow}).
The empty square with the label $8$ in (c) represents the initial
value of the average eight-point vertex given $\propto J^{2}$. Note,
that the flow equation (a) for the self-energy is exact, while on the right-hand side of
the flow equations (b) and (c) for the four-point and the six-point vertices
we have retained only the contributions from the vertices
with the largest number of legs, which give the leading contributions for large $N$.}
\label{fig:flowtrunc}
\end{figure}
At $T=0$ and $\mu=\mu_{*}\left(0\right)\approx0.212J$ the density jumps from $n\approx 0.76$ to $1$. For $\mu=-\mu_*(0)$ it jumps from $0.24$ to $0$. At $T>0$, $\mu_{*}$
terminates at a critical end-point. In Fig.~1 we reproduce
the phase diagram already discussed in Ref.~\cite{Azeyanagi2018}.

\emph{Functional renormalization group:} Identifying the critical
surface and the relevant scaling variables should give important additional
insights into a model with scale-invariant solutions. Surprisingly,
in spite of significant recent attention to the SYK model, we have
not been able to find an answer to these fundamental issues in the
published literature. In order to derive the FRG flow equations we
perform the average over disorder configurations which yields, in the absence of
replica symmetry breaking,  an effective four-body interaction with Euclidean action
\begin{eqnarray}
S_{{\rm 8}} & = & -\frac{J^{2}}{N^{3}}\frac{1}{(2!)^{2}}\int_{0}^{\beta}d\tau\int_{0}^{\beta}d\tau^{\prime}\left|\sum_{i}\bar{c}_{i}(\tau)c_{i}(\tau^{\prime})\right|^{4},\label{eq:S8dis}
\end{eqnarray}
where $c_i ( \tau )$ and $\bar{c}_i ( \tau )$ are Grassmann variables and $\beta =1/T$.
The FRG is then based on introducing a frequency-dependent regulator
in the Gaussian part of the action, governed by the cutoff-parameter
$\Lambda$ which suppresses excitations with energy below $\Lambda$.
The physical problem is then recovered in the limit $\Lambda\rightarrow0$.
For our purpose  it suffices to work with a sharp frequency cutoff
and use the scale-dependent bare propagator $G_{0,\Lambda}(\omega)=\Theta(|\omega|-\Lambda)/\left(i\omega+\mu\right)$.
Following the usual procedure~\cite{Salmhofer01,Kopietz01,Kopietz10,Metzner12,Dupuis20},
we can now derive the Wetterich equation~\cite{Wetterich93} for the
scale-dependent effective action $\Gamma_{\Lambda}[ \langle \bar{c} \rangle , \langle c \rangle]$,
which generates an infinite hierarchy of coupled integro-differential
equations for the scale-dependent one-particle irreducible $k$-point vertices $\Gamma_{\Lambda}^{\left(k\right)}$.
This infinite hierarchy simplifies considerably in the large-$N$
limit. To leading order in $1/N$, it is sufficient to  keep in the flow equation for the
$k$-point vertex only diagrams involving the $(k+2)$-point vertex, 
see Fig.~\ref{fig:flowtrunc}. For the flow equation of the self-energy
$\Sigma_{\Lambda}(\omega)$ ($k=2$) this is exact, while for the 
higher-order vertices the neglected diagrams are sub-leading in $1/N$.
The bare four-body interaction (\ref{eq:S8dis}) implies that
at the initial scale only the antisymmetrized  $8$-point vertex 
$\Gamma_{\Lambda_{0}}^{\left(8\right)}\propto-J^{2}$ is finite.
To proceed it is convenient
to parametrize the scale-dependence of the couplings in terms of the
logarithmic flow parameter $l=\ln(\Lambda_{0}/\Lambda)$, where $\Lambda_0 \gg J$ and define
the dimensionless rescaled couplings $r_{l}=Z_{\Lambda}^{2}\left(\mu-\Sigma_{\Lambda}\left(0\right)\right)^{2}/\Lambda^{2}$,
$u_{l}=Z_{\Lambda}^{2}\Gamma_{\Lambda}^{(4)}(0)/\Lambda$, and 
$g_{l}=J^{2}Z_{\Lambda}^{4}/\Lambda^{2}$
which is proportional to the rescaled  six-point vertex at vanishing frequencies.
The wave-function renormalization $Z_{\Lambda}$  is obtained
from the low-frequency expansion of the scale-dependent self-energy,
$\Sigma_{\Lambda}(\omega)=\Sigma_{\Lambda}(0)+(1-Z_{\Lambda}^{-1})i\omega+\cdots$, and
defines the scale-dependent anomalous dimension $\eta_{\Lambda}=\Lambda\partial_{\Lambda}\ln Z_{\Lambda}.$
Using $-\Lambda\partial_{\Lambda}=\partial_{l}$ the truncation illustrated
in Fig.~\ref{fig:flowtrunc} yields the flow equations~\cite{Smit20}
 \begin{subequations}
\label{eq:gflow}
\begin{eqnarray}
\partial_{l}r_{l} & = & 2(1-\eta_{l})r_{l}-\frac{2}{\pi}\frac{r_{l}u_{l}}{1+r_{l}}, \\
\partial_{l}u_{l} & = & (1-2\eta_{l})u_{l}+\frac{4}{\pi}\frac{g_{l}}{1+r_{l}}\left[\frac{1}{1+r_{l}}-\frac{3}{4}\right], \\
\partial_{l}g_{l} & = & 2(1-2\eta_{l})g_{l}.
\end{eqnarray}
 \end{subequations}
Analyzing Eq.(\ref{eq:gflow}) we find the fixed points shown in Fig.~\ref{fig:flow3d}:
i) the unstable vacuum fixed point $V$ at $r=u=g=0$; ii) the discontinuity
fixed point $D$ with one relevant direction; iii) the stable NFL
fixed point $S_{+}$ with $\Sigma\left(0\right)=\mu$, preserving
scaling. Here, the renormalized two-body interaction diverges to $+\infty$
and the coupling $g$ approaches $\pi^{2}/3$; and iv) the stable
integer-valence fixed point $S_{-}$ with $\Sigma\left(0\right)\neq\mu$,
corresponding to a gapped state yet with a divergent attractive two-body
interaction $u\rightarrow-\infty$. Notice, the large-$N$ self-consistency equations
 for the self-energy $\Sigma(\omega)$ can be recovered from
our FRG approach. This is achieved by substituting our result for
the four-point vertex into the flow equation for the self-energy and
integrating over the flow parameter $\Lambda$.  
\begin{figure}[tb]
\begin{centering}
\includegraphics[width=0.4\textwidth]{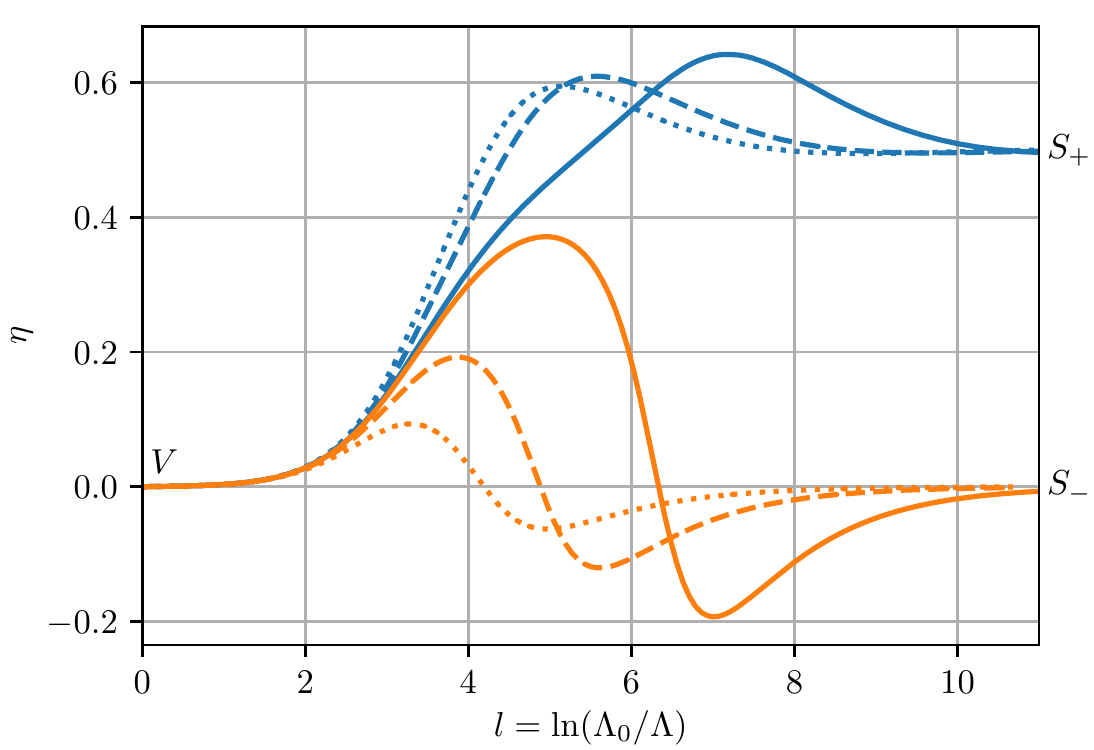} \\
\includegraphics[scale=0.63]{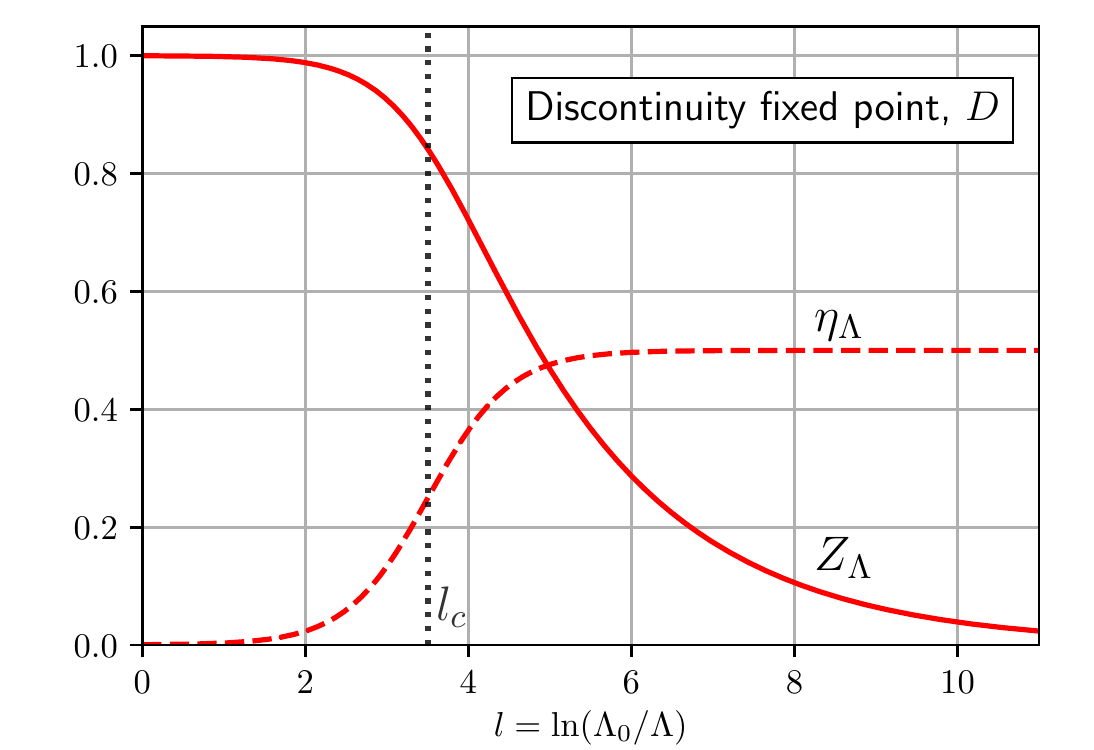} 
\par\end{centering}
\caption{\textit{Top}: RG flow of the anomalous dimension for different initial
conditions near the vacuum fixed point. The NFL sink $S_{+}$ yields
$\eta=1/2$ (blue curves), while the orange curves flow into the integer-valence
fixed point $S_{-}$ where the system is gapped and $\eta=0$.\textit{
Bottom}: flow of the wave-function renormalization $Z_{\Lambda}$
and the scale-dependent anomalous dimension $\eta$ right at the discontinuity
fixed point $D$. The dashed line marks the crossover scale $l_{c}=\ln(\Lambda_{0}/\Lambda_{c})$
determined by Eq.~(\ref{eq:Zflow}).}
\label{fig:flowetag} 
\end{figure}

In order to characterize the fermionic dynamics we determine the scale-dependent anomalous dimension via
\begin{equation}
\eta_{l}=\frac{\Lambda Z_{\Lambda}}{\beta}\sum_{\omega_{2}}\dot{G}_{\Lambda}(\omega_{2})\left.\frac{\partial\Gamma_{\Lambda}^{(4)}(\omega_{1},\omega_{2};\omega_{2},\omega_{1})}{\partial(i\omega_{1})}\right|_{\omega_{1}=0},\label{eq:eta}
\end{equation}
with single-scale propagator $\dot{G}_\Lambda(\omega)$~\cite{Kopietz10,Metzner12}. The four-point vertex 
$\Gamma_{\Lambda}^{(4)}$ is expressed in terms of scale dependent
particle-particle and particle-hole susceptibilities~\cite{Smit20}.
The behavior of $\eta_{l}$ during the flow from the vicinity of the
vacuum fixed point with $\eta=0$ is shown in Fig.~\ref{fig:flowetag}
for slightly different initial conditions. As expected, the NFL fixed
point $S_{+}$ has $\eta=1/2$ in agreement with Eq.~(\ref{eq:NFLsol}),
while we find $\eta =0$ for the integer-valence fixed point $S_{-}$,
consistent with Eq.~(\ref{eq:IVsol}). The most remarkable finding of
our analysis is however that $\eta=1/2$ at the discontinuity fixed
point, see lower part of Fig.~\ref{fig:flowetag}. While this fixed
point describes a discontinuous change of the particle number, the
fermionic dynamics is  quantum critical. The first order
transition is characterized by critical slowing down of the fermions.
A direct implication would be the scaling behavior of the longitudinal
dielectric constant $\varepsilon_{l}\left(\omega\right)\propto\omega^{2\eta-3}$
right at phase coexistence. We followed Ref.~\cite{Song17} to
determine the electromagnetic response. Power law behavior  of $\varepsilon_{l}\left(\omega\right)$ that would imply $\eta \approx 0.75$  was recently observed in Refs.~\cite{Pustogow19,Rosslhuber20} at a
Mott transition. 
While there is no reason to expect that the SYK model yields the value of the measured exponent, its observation supports the concept of phase coexistence as a genuine critical  state of matter.

\emph{Discontinuity fixed point:} At the discontinuity fixed point
$D$ all rescaled couplings approach finite limits. This enables us
to analyze the linearized flow in its vicinity. Using   $\eta=1/2$ we can determine the
numerical values of our couplings by demanding that all scale derivatives
in Eq.(\ref{eq:gflow}) vanish. We obtain 
\begin{equation}
r_{\ast}=\frac{1}{3},\;\;\;u_{\ast}=\frac{2\pi}{3},\;\;\;g_{\ast}=\frac{1}{2c}\approx8.52.\;\;\;\label{eq:fixedpointvalues}
\end{equation}
The result for $g_{*}$ follows from an analysis of Eq.(\ref{eq:eta})
near $D$ which yields $\eta_{\Lambda}=cg_{\Lambda}$ with numerical
coefficient $c=3\left(\log2-\frac{1}{2}\right)/\pi^{2}\approx0.0587$. The precise value of $c$ may weakly depend on details of the regularization scheme. 
From Fig.~\ref{fig:flow3d} we see that the discontinuity fixed point
has an unstable and a stable direction. To calculate the corresponding
scaling variables, we linearize the flow in its vicinity. Setting
$r_{l}=r_{\ast}+\delta r_{l}$, and similar for $u_{l}$ and $g_{l}$,
we obtain the linearized flow equations. Within our truncation $\delta g_{l}$
decouples  and we can focus on
the $\delta r$-$\delta u$-plane. 
With $\delta u_{l}=\frac{3}{4}\sqrt{\frac{3}{c}}\delta y_{l}$
we obtain
\begin{eqnarray}
\partial_{l}\left(\begin{array}{c}
\delta r_{l}\\
\delta y_{l}
\end{array}\right)=\left(\begin{array}{cc}
\frac{1}{4} & -a\\
-a & 0
\end{array}\right)\left(\begin{array}{c}
\delta r_{l}\\
\delta y_{l}
\end{array}\right),\label{eq:linflow}
\end{eqnarray}
with numerical constant $a=\frac{3}{8\pi}\sqrt{{3}/{c}}.$ The
eigenvalues of the matrix are $\lambda_{+}\approx0.987$ and $\lambda_{-}=1/4-\lambda_{+}.$
The corresponding eigen-directions in the $r$-$u$-plane follow easily.
Near the transition, the relevant eigenvector is proportional to $\mu-\mu_{*}$.
Hence, near the transition, the grand canonical potential per fermion
species should - in addition to a regular piece - be characterized
by a singular contribution with following scaling behavior,
\begin{equation}
\Omega\left(\mu\right)=\Omega_{{\rm reg}}\left(\mu\right)+e^{-l}\Omega_{{\rm sing}}\left(e^{\lambda_{+}l}\left(\mu-\mu_{*}\right)\right).
\end{equation}
This yields for the particle density $n=-\partial\Omega/\partial\mu=n_{{\rm reg}}\pm B_{\pm}\left|\mu-\mu_{*}\right|^{-1 + 1/\lambda_{+}}$. Keeping in mind that our determination of $c$ is approximate, the above
result for $\lambda_{+}$ is consistent with $\lambda_{+}=1$, yielding
indeed a discontinuity of $n$ at $\mu_{*}$. According to scaling
arguments near first order transitions~\cite{Nienhuis75,Fisher82}
the relevant eigenvalue is the space dimension, i.e. $\lambda_{+}=d$.
For our zero-dimensional quantum system with one time direction this
yields indeed $\lambda_{+}=1$.

The linear relation between $\eta_{\Lambda}$ and $g_{\Lambda}$ near
the discontinuity fixed point can also be used to determine the crossover
energy to the regime with $\eta\approx\frac{1}{2}$. Expressing $g_{\Lambda}$
and $\eta_{\Lambda}$ in terms of $Z_{\Lambda}$, we obtain the closed
flow equation of the wave-function renormalization near  $D$,
\begin{equation}
\partial_{\Lambda}Z_{\Lambda}=cJ^{2}\Lambda^{-3} Z_{\Lambda}^{5}.\label{eq:Zflow}
\end{equation}
From the exact solution of this equation with initial condition $Z_{\Lambda_{0}}=1$
we find that critical behavior with $\eta\rightarrow 1/2$ sets in at the crossover scale $\Lambda_{c}=\sqrt{2c}J\approx0.343J$, see the lower part of Fig.~\ref{fig:flowetag}.

\emph{Integer-valence fixed point:} Finally, we elucidate our finding
that the rescaled two-body interaction $u_{l}$ at the integer valence fixed point  $S_{-}$ 
shown in Fig.~\ref{fig:flow3d}  flows to $-\infty$, which corresponds to a finite attractive
two-body interaction. Of course,  the
single-particle properties  for occupations $n=0$ or $1$  are trivial.  A single electron or hole added to the system has no particle
to interact with.  This does not apply to higher-particle correlations.  Following the classical work of Galitskii~\cite{Galitskii58}
the two-particle vertex of a such a dilute system can be obtained by summing up
all particle-particle ladder diagrams. We perform this analysis prior to the disorder averaging,
\begin{equation}
\Gamma_{ijkl}\left(\omega\right)=J_{ijkl}-\sum_{m<n}J_{ijmn}\Gamma_{mnkl}\left(\omega\right)\chi\left(\omega\right),\label{eq:LS compact}
\end{equation}
with particle-particle bubble $\chi\left(\tau\right)=G^{2}\left(\tau\right)$.
Summing up the series and averaging term-by-term over the $J_{ijkl}$
 yields at large $N$ for $i<j$ and $k<l$ and after analytic
continuation to retarded functions, 
\begin{equation}
\text{\ensuremath{\Gamma^{\rm ret}_{ijkl}\left(\epsilon-2\mu\right)}}=-\delta_{ik}\delta_{jl}\epsilon\left(1-\frac{2}{1+\sqrt{\frac{\epsilon^{2}-4J^{2}/N}{\epsilon^{2}}}}\right).
\end{equation}
This result implies that a spectrum of bonding and anti-bonding two-particle
scattering states of bandwidth $4J/\sqrt{N}$ emerges through ${\rm Im}\Gamma_{ijkl}\neq0$
below and above the trivial two-particle energy $-2\mu$. While the
single-particle physics of the integer-valence phase is trivial, pairs
of correlated electrons or holes propagate coherently, a behavior
that is signaled by $u\rightarrow-\infty$ in Fig.~\ref{fig:flow3d}.

In summary, we formulated the functional RG of the SYK model and determined
its global RG flow. Our approach reproduces known results of the two
stable phases, the non-Fermi liquid state and the integer-valence
state, see Fig.\ref{fig:phasediagram}. For the integer-valence phase
it also revealed interesting two-particle correlations. Most importantly,
the  FRG allows for insights into the discontinuity fixed
point that separates these two phases. The relevant scaling dimension
of this fixed point is consistent with the behavior near first-order
transitions~\cite{Nienhuis75,Fisher82}. In addition, fermionic
excitations at phase coexistence behave quantum critical. Such a behavior cannot
be captured in terms of a local Ising variable.
Thus, quantum first-order transitions that do not break a symmetry
are shown to be dramatically altered by the presence of non-Fermi
liquid electronic excitations.

\emph{Acknowledgements:} This work was supported by the Deutsche Forschungsgemeinschaft
(DFG, German Research Foundation) - TRR 288 - 422213477 (project A07).
We are grateful to M. A. Axenovich and M. Garst for helpful discussions.

\end{document}